\documentclass[preprint2]{aastex62}
\renewcommand\plotone[1]{\includegraphics[width=\linewidth]{#1}}

\shorttitle{Dwarf Carbon Stars}
\shortauthors{Harris et al.}

\begin{document}

\title{Distances of Dwarf Carbon Stars}

\author{Hugh C. Harris}
\affiliation{US Naval Observatory, Flagstaff Station, 10391 W. Naval Observatory
Road, Flagstaff, AZ 86005-8521, USA}

\author{Conard C. Dahn}
\affiliation{US Naval Observatory, Flagstaff Station, 10391 W. Naval Observatory
Road, Flagstaff, AZ 86005-8521, USA}

\author{John P. Subasavage}
\affiliation{US Naval Observatory, Flagstaff Station, 10391 W. Naval Observatory
Road, Flagstaff, AZ 86005-8521, USA}

\author{Jeffrey A. Munn}
\affiliation{US Naval Observatory, Flagstaff Station, 10391 W. Naval Observatory
Road, Flagstaff, AZ 86005-8521, USA}

\author{Blaise J. Canzian}
\affiliation{L-3 Communications/Brashear, 615 Epsilon Dr., Pittsburgh, PA 15238-2807, USA}

\author{Stephen E. Levine}
\affiliation{Lowell Observatory, 1400 W. Mars Hill Road, Flagstaff, AZ  86001-4499, USA}

\author{Alice B. Monet}
\affiliation{US Naval Observatory, Flagstaff Station, 10391 W. Naval Observatory
Road, Flagstaff, AZ 86005-8521, USA}

\author{Jeffrey R. Pier}
\affiliation{US Naval Observatory, Flagstaff Station, 10391 W. Naval Observatory
Road, Flagstaff, AZ 86005-8521, USA}

\author{Ronald C. Stone}
\altaffiliation{Deceased}
\affiliation{US Naval Observatory, Flagstaff Station, 10391 W. Naval Observatory
Road, Flagstaff, AZ 86005-8521, USA}

\author{Trudy M. Tilleman}
\affiliation{US Naval Observatory, Flagstaff Station, 10391 W. Naval Observatory
Road, Flagstaff, AZ 86005-8521, USA}

\author{William I. Hartkopf}
\affiliation{US Naval Observatory, 3450 Massachusetts Ave. N. W.,
  Washington, DC 20392-5420, USA}

\correspondingauthor{Munn, J. A.}
\email{jam@nofs.navy.mil}

\begin{abstract}
Parallaxes are presented for a sample of 20 nearby dwarf
carbon stars.  \added{The inferred luminosities cover almost two orders of magnitude}.  Their absolute magnitudes and tangential
velocities confirm prior expectations that some originate
in the Galactic disk, although more than half of this sample
are halo stars.  Three stars are found to be \added{astrometric} binaries,
and orbital elements are determined; their semimajor axes
are 1--3 AU, consistent with the size of an AGB mass-transfer
donor star.
\end{abstract}


\section{Introduction}
Distances, absolute magnitudes, and luminosities for the
dwarf carbon (dC) stars are uncertain.  They tend to be faint,
and at sufficiently large distances, that parallaxes are not
easy to measure.  To date only three have published parallaxes
\citep{1977ApJ...216..757D,1998ApJ...502..437H};  those three have similar colors and absolute
magnitudes.  However, properties of the many dwarf carbon stars
discovered in recent years \citep{2002AJ....124.1651M, 2004AJ....127.2838D,
2013ApJ...765...12G}
suggest that they are likely to have a broad range of
physical properties, and expectations about their origin
\citep{1994ApJ...423..723G} suggest that they are likely to be produced in
both disk and halo populations with a range of metallicities and
absolute magnitudes.
This paper presents parallaxes for an expanded sample of
dwarf carbon stars, with a goal of investigating the range
of absolute magnitudes over which these stars extend.

\section{Data}
\subsection{Sample}
The sample of 20 dC stars in this paper consists of 13 targets taken
from the Sloan Digital Sky Survey \citep[SDSS;][]{1996AJ....111.1748F,
1998AJ....116.3040G,2006AJ....131.2332G,2000AJ....120.1579Y} and seven targets from
other papers in the literature.  Three have parallaxes published
previously \citep{1998ApJ...502..437H}, and here we give an improved
parallax for two of them (LHS 1075 and CLS 96), and an entirely new
parallax for the third (G77-61).  These targets were chosen to be
definite dwarfs (as opposed to giant carbon stars) based on their
significant proper motions.  They were selected to be ``nearby'' (with
distances likely to be within 300 pc) so as to have significant
parallaxes; criteria used to select the sample included large proper
motion, bright apparent magnitude for a candidate's color, and a range
of colors and band strengths so as to include a variety of types of
stars.  The sample does not include any of the warm, CH-like stars
from the SDSS, because the estimated distances for even the brightest
were $> 500\ {\rm pc}$, too distant to get a meaningful parallax
measurement.

\subsection{Astrometry}
Images have been taken with the U.S. Naval Observatory's (USNO) 1.55 m Strand
Astrometric Reflecting Telescope in Flagstaff, AZ.  Descriptions of
the telescope, the CCD cameras and filters used here, and the observing and
processing procedures are given by \citet{1992AJ....103..638M} and
\citet{2017AJ....154..147D}.  The astrometric results are given in
Table~\ref{tab:astrometry}.
Column (1) gives the identifying number in the
2MASS Point Source Catalog \citep{2006AJ....131.1163S}.  The 2MASS J number
provides an unambiguous link to SIMBAD where many alternate names and much
additional information can be found.  Column (2) gives alternate names, usually
the name by which the star was first identified as a carbon star.  Columns (5)
and (6) indicate the camera and filter employed for each parallax
determination, while
Columns (7), (8), and (9) give the number of acceptable CCD frames
(observations), the number of separate nights on which those observations were
obtained, and the number of reference stars employed in astrometric solutions,
respectively.  Columns (10) and (11) give the years observed and total epoch
range, respectively.
The derived relative parallax and its mean (standard) uncertainty are given in
Column (12) (throughout this work we refer to the trigonometric parallax angle
as $\pi$).  The relative total proper motion and its uncertainty follow in
Column (13), and the position angle of the proper motion is given in Column
(14).  The error in the position angle reflects the uncertainty in the
orientation of the CCD in its dewar, and of the dewar on the telescope.  The
derived absolute parallax and its uncertainty are presented in Column (15), and
the calculated tangential velocities and their uncertainties
are given in Column (16).

Many of the 20 stars in this paper are
near the faint limit for this observing program and require long
exposure times and good seeing.  As a consequence, we have obtained
fewer observations than desired, some with lower signal-to-noise ratio than
desired, and the resulting parallax values have larger errors than
desired.  Nevertheless, the parallaxes are significant for all stars:
the fractional error in parallax has a median value of 9\%, and ranges
from 3\% in the best case to 23\% in the worst case.

\begin{splitdeluxetable*}{ccccccrrrBcrr@{ $\pm$ }rr@{ $\pm$ }rr@{ $\pm$ }rr@{ $\pm$ }rr@{ $\pm$ }rcc}
\voffset-10pt{}
\tabletypesize{\tiny}
\tablewidth{0pt}
\tablecaption{Astrometric Results
\label{tab:astrometry}}
\tablehead{
  \colhead{}  &
  \colhead{}  &
  \colhead{R.A.\tablenotemark{a}}  &
  \colhead{Decl.\tablenotemark{a}}  &
  \colhead{}  &
  \colhead{}  &
  \colhead{}  &
  \colhead{}  &
  \colhead{}  &
  \colhead{}  &
  \colhead{$\Delta$T} & 
  \multicolumn{2}{c}{$\pi_{\rm rel}$} &
  \multicolumn{2}{c}{$\mu_{\rm rel}$} &
  \multicolumn{2}{c}{P.A.} &
  \multicolumn{2}{c}{$\pi_{\rm abs}$} &
  \multicolumn{2}{c}{$V_{\rm tan}$} &
  \colhead{} &
  \colhead{} \\
  \colhead{2MASS J} & 
  \colhead{Alternate Names} & 
  \multicolumn{2}{c}{(J2000.0)} &
  \colhead{Cam.} & 
  \colhead{Filt.} & 
  \colhead{Nf} & 
  \colhead{Nn} & 
  \colhead{Ns} & 
  \colhead{Coverage}  &
  \colhead{(yr)} &
  \multicolumn{2}{c}{(mas)} &
  \multicolumn{2}{c}{(mas yr$^{\rm -1}$)} &
  \multicolumn{2}{c}{(deg)} &
  \multicolumn{2}{c}{(mas)} &
  \multicolumn{2}{c}{(km s$^{\rm -1}$)} &
  \colhead{Pop.\tablenotemark{b}} &
  \colhead{Notes\tablenotemark{c}} \\
  \colhead{(1)} &
  \colhead{(2)} &
  \colhead{(3)} &
  \colhead{(4)} &
  \colhead{(5)} &
  \colhead{(6)} &
  \colhead{(7)} &
  \colhead{(8)}  &
  \colhead{(9)}  &
  \colhead{(10)}  &
  \colhead{(11)}  &
  \multicolumn{2}{c}{(12)} &
  \multicolumn{2}{c}{(13)} &
  \multicolumn{2}{c}{(14)} &
  \multicolumn{2}{c}{(15)} &
  \multicolumn{2}{c}{(16)} &
  \colhead{(17)} &
  \colhead{(18)}}

\startdata
00260048$-$1918519 & LHS1075 (LP765-18)        & 00 26 00.2  & $-$19 18 52  &
TEK2K  & A2$-$1  &  148 & 129 &   7 &  1992.76$-$2002.62 &   9.86 &   6.26 & 0.63
&  613.5 & 0.2 &  180.8 & 0.2 &   7.33 & 0.63 &  397 & 34 & 2 \\
01202853$-$0836307 & SDSS J012028.55-083630.8  & 01 20 28.5  & $-$08 36 31  &
EEV24  &  I$-$2  &  112 &  92 &  10 &  2008.72$-$2017.00 &   8.28 &   1.22 & 0.47
&  148.9 & 0.1 &  108.5 & 0.2 &   2.32 & 0.48 &  304 & 63 & 4 &   1 \\
01215031$+$0113024 & LP587-45 (NLTT4523)       & 01 21 50.3  & $+$01 13 03  &
TEK2K  &  I$-$2  &  145 & 144 &   8 &  2003.72$-$2015.64 &  11.92 &   2.64 & 0.34
&  230.3 & 0.1 &  124.8 & 0.2 &   3.84 & 0.35 &  284 & 26  & 2   \\
03323808$+$0157599 & G77-61 (LHS1555)          & 03 32 38.1  & $+$01 58 00  &
EEV24  & A2$-$1  &  117 &  71 &  13 &  2010.75$-$2017.10 &   6.35 &  11.53 & 0.56
&  765.7 & 0.2 &  166.3 & 0.2 &  12.77 & 0.56 &  284 & 12 &  2 & 2 \\
07425720$+$4659186 & LSPM J0742+4659           & 07 42 57.2  & $+$46 59 18  &
EEV24  &  I$-$2  &  277 & 145 &  12 &  2008.88$-$2017.10 &   8.22 &   5.95 & 0.15
&  153.2 & 0.1 &  205.1 & 0.2 &   6.76 & 0.16 &  107 & 3 &  4 &  3 \\
08180742$+$2234290 & SDSS J081807.45+223427.6  & 08 18 07.4  & $+$22 34 29  &
EEV24  &  I$-$2  &  127 & 106 &  20 &  2008.29$-$2017.25 &   8.96 &   1.66 & 0.26
&  237.7 & 0.1 &  168.7 & 0.2 &   2.47 & 0.27 &  456 & 50 & 4 \\
08345114$+$0740088 & SDSS J083451.15+074008.8  & 08 34 51.2  & $+$07 40 09  &
EEV24  & A2$-$1  &   37 &  30 &  17 &  2012.94$-$2017.10 &   4.16 &   1.80 & 0.59
&   66.5 & 0.3 &  123.9 & 0.3 &   2.86 & 0.59 &  110 & 23 & 1 \\
09332463$-$0031445 & HE 0930-0018              & 09 33 24.6  & $-$00 31 45  &
TEK2K  & A2$-$1  &  100 &  97 &  14 &  2004.05$-$2010.30 &   6.25 &   5.76 & 0.45
&   62.6 & 0.2 &  299.6 & 0.3 &   6.70 & 0.46 &   44 & 3 & 1 \\
10542941$+$3402259 & CLS31                     & 10 54 29.4  & $+$34 02 26  &
EEV24  &  I$-$2  &   63 &  61 &   9 &  2008.29$-$2015.31 &   7.02 &   1.00 & 0.39
&  120.1 & 0.1 &  205.7 & 0.2 &   1.66 & 0.39 &  343 & 81 & 2 \\
11203487$+$5559373 & SDSS J112034.86+555937.2  & 11 20 34.9  & $+$55 59 37  &
TEK2K  &  I$-$2  &  204 & 186 &   7 &  2003.18$-$2017.31 &  14.13 &   0.83 & 0.30
&   87.7 & 0.1 &  203.2 & 0.2 &   1.77 & 0.31 &  243 & 45 &  2 & 1 \\
13533300$-$0040395 & SDSS J135333.01-004039.4  & 13 53 33.1  & $-$00 40 40  &
TEK2K  &  I$-$2  &  222 & 198 &  11 &  2003.27$-$2016.45 &  13.18 &   1.29 & 0.26
&   46.3 & 0.1 &  240.0 & 0.2 &   2.27 & 0.27 &  100 & 13 & 1 \\
14531880$+$6004209 & SDSS J145318.83+600421.0  & 14 53 18.8  & $+$60 04 21  &
EEV24  &  I$-$2  &  111 &  93 &   9 &  2008.34$-$2017.49 &   9.15 &   3.34 & 0.46
&  224.3 & 0.1 &  267.2 & 0.2 &   4.51 & 0.47 &  236 & 25 & 2 \\
14572597$+$2341257 & SDSS J145725.86+234125.5  & 14 57 26.0  & $+$23 41 26  &
EEV24  &  I$-$2  &   44 &  41 &  21 &  2012.46$-$2017.49 &   5.03 &   6.76 & 0.47
&  360.9 & 0.2 &  259.9 & 0.2 &   7.65 & 0.47 &  224 & 14 &  2 & 4 \\
15270276$+$4345172 & SDSS J152702.74+434517.3  & 15 27 02.8  & $+$43 45 17  &
TEK2K  &  I$-$2  &  287 & 231 &  15 &  2003.19$-$2017.46 &  14.27 &   2.51 & 0.32
&   41.1 & 0.1 &  308.9 & 0.2 &   3.53 & 0.32 &   97 & 5 & 1 \\
15480927$+$3227247 & SDSS J154809.22+322724.9  & 15 48 09.2  & $+$32 27 25  &
TEK2K  &  I$-$2  &  115 &  96 &  15 &  2008.25$-$2015.56 &   7.31 &   2.78 & 0.39
&  141.7 & 0.1 &  291.4 & 0.2 &   3.98 & 0.40 &  169 & 17 & 3 \\
15523734$+$2927591 & LP328-57 (CLS96)          & 15 52 37.4  & $+$29 27 59  &
TEK2K  & A2$-$1  &  242 & 183 &  10 &  1992.20$-$2003.35 &  11.15 &   3.45 & 0.41
&  281.1 & 0.1 &  227.6 & 0.2 &   4.70 & 0.43 &  284 & 26 & 2 \\
16223288$+$4237538 & LP225-12 (NLTT42660)      & 16 22 32.9  & $+$42 37 54  &
TEK2K  &  I$-$2  &  293 & 231 &   8 &  2003.48$-$2017.57 &  14.09 &   7.50 & 0.20
&  187.4 & 0.1 &  325.8 & 0.2 &   8.99 & 0.24 &  101 & 5  &  4 & 3 \\
16313278$+$3553285 & LP275-54 (NLTT43012)      & 16 31 32.8  & $+$35 53 29  &
EEV24  &  I$-$2  &  147 & 106 &  17 &  2008.29$-$2015.54 &   7.25 &   1.57 & 0.43
&  186.1 & 0.1 &  308.7 & 0.2 &   2.51 & 0.43 &  583 & 60 & 2 \\
21051653$+$2514486 & LSR J2105+2514            & 21 05 16.5  & $+$25 14 49  &
TEK2K  &  I$-$2  &  165 & 146 &  15 &  2003.50$-$2010.72 &   7.22 &   9.09 & 0.27
&  560.4 & 0.1 &  150.7 & 0.2 &   9.81 & 0.27 &  271 & 7 & 2 \\
21493784$-$1138285 & LP758-43 (NLTT52182)      & 21 49 37.8  & $-$11 38 29  &
TEK2K  & A2$-$1  &  277 & 227 &   6 &  2003.56$-$2017.97 &  14.41 &   6.18 & 0.32
&  252.2 & 0.1 &   89.8 & 0.2 &   7.20 & 0.33 &  166 & 8  &   3 & 3 \\
\enddata
\tablecomments{Table~\ref{tab:astrometry} is published in machine-readable format.} 
\tablenotetext{a}{Coordinates are from the 2MASS catalog.}
\tablenotetext{b}{Estimated Galactic population membership (see Section~\ref{sec:discussion}):  (1) disk, (2) halo, (3) intermediate, (4) contradictory.}
\tablenotetext{c}{Notes on individual objects. (1) Hints of a possible very low
  amplitude perturbation not confirmed. (2) No evidence seen corresponding to
  the 245 day radial velocity period reported by \citet{1986ApJ...300..314D}.
  (3) Tabulated astrometry after removal of the perturbation discussed in
  Section~\ref{sec:orbits} below.
  (4) This is the common proper motion companion to 2MASS J14572616+2341227
  (LSPM J1457+2341S)
  for which $\pi_{\rm abs} = 5.77 \pm 1.48\ {\rm mas}$ was reported in \citet{2017AJ....154..147D}.
  \replaced{A}{These a}dditional observations further support the physical nature of the pair.
  The weighted mean astrometric results for the pair are reported above.
  \citet{2016ApJS..224...36K} reported evidence for a third component
  based on the ALLWISE motion survey.}
\end{splitdeluxetable*}

\subsection{Photometry}
Photometry of most of the target
stars and their surrounding reference stars was obtained with the USNO
1.0 m and 1.55 m telescopes using Johnson-Cousins $BVI$ filters.  These data were
used in the astrometric processing above to correct for differential
color refraction (DCR).  For those fields lacking $BVI$ photometry, $gri$ data
from the SDSS database were transformed to $BVI$ using the relations of \citet{2007AJ....134..973I} and then used for the DCR corrections.  Photometric data for the 20 dC target stars is given in
Table~\ref{tab:photometry}
including $JHK_S$ from 2MASS.

\begin{deluxetable*}{lr@{ $\pm$ }rr@{ $\pm$ }rr@{ $\pm$ }rcr@{ $\pm$ }rr@{
      $\pm$ }rr@{ $\pm$ }r}
\tabletypesize{\small}
\tablewidth{0pt}
\tablecaption{Photometric Results
\label{tab:photometry}}
\tablehead{
	\colhead{2MASS J} & 
	\multicolumn{2}{c}{$B$} &
	\multicolumn{2}{c}{$V$} &
	\multicolumn{2}{c}{$I$} &
        \colhead{Notes\tablenotemark{a}} &
	\multicolumn{2}{c}{$J$} &
	\multicolumn{2}{c}{$H$} &
	\multicolumn{2}{c}{$K_S$} \\
	\colhead{(1)}  &
	\multicolumn{2}{c}{(2)}  &
	\multicolumn{2}{c}{(3)}  &
	\multicolumn{2}{c}{(4)}  &
	\colhead{(5)}  &
	\multicolumn{2}{c}{(6)}  &
	\multicolumn{2}{c}{(7)}  &
	\multicolumn{2}{c}{(8)}}
\startdata
00260048$-$1918519 & 16.76   & 0.05    & 15.07 & 0.03 & 13.46 & 0.03 & 1 & 12.540 & 0.024 & 11.918 & 0.026 & 11.587 & 0.027 \\
01202853$-$0836307 & 19.62   & 0.05    & 17.72 & 0.03 & 15.82 & 0.03 & 1 & 14.990 & 0.043 & 14.224 & 0.048 & 13.665 & 0.047 \\
01215031$+$0113024 & 19.65   & 0.05    & 17.76 & 0.03 & 16.08 & 0.03 & 1 & 15.107 & 0.049 & 14.251 & 0.045 & 13.805 & 0.045 \\
03323808$+$0157599 & 15.64   & 0.01    & 13.89 & 0.01 & 12.33 & 0.01 & 2 & 11.470 & 0.022 & 10.844 & 0.024 & 10.480 & 0.021 \\
07425720$+$4659186 & 18.75   & 0.05    & 16.65 & 0.03 & 14.89 & 0.03 & 1 & 13.520 & 0.025 & 12.681 & 0.026 & 12.182 & 0.026 \\
08180742$+$2234290 & 17.90   & 0.15    & 16.24 & 0.10 & 14.60 & 0.10 & 3 & 13.744 & 0.026 & 12.990 & 0.026 & 12.634 & 0.025 \\
08345114$+$0740088 & 19.40   & 0.15    & 17.28 & 0.10 & 15.48 & 0.10 & 3 & 14.416 & 0.034 & 13.744 & 0.040 & 13.426 & 0.029 \\
09332463$-$0031445 & 16.12   & 0.05    & 14.64 & 0.03 & 13.14 & 0.03 & 1 & 12.190 & 0.023 & 11.546 & 0.024 & 11.335 & 0.023 \\
10542941$+$3402259 & 19.26   & 0.15    & 17.79 & 0.10 & 16.30 & 0.10 & 3 & 15.501 & 0.051 & 15.018 & 0.065 & 14.655 & 0.082 \\
11203487$+$5559373 & 18.20   & 0.05    & 16.66 & 0.03 & 15.25 & 0.03 & 1 & 14.408 & 0.031 & 13.771 & 0.043 & 13.633 & 0.043 \\
13533300$-$0040395 & \nodata & \nodata & 17.29 & 0.03 & 15.60 & 0.03 & 1 & 14.609 & 0.036 & 13.798 & 0.036 & 13.609 & 0.051 \\
14531880$+$6004209 & 19.95   & 0.15    & 18.29 & 0.10 & 16.49 & 0.10 & 3 & 15.609 & 0.054 & 15.053 & 0.071 & 14.764 & 0.092 \\
14572597$+$2341257 & 19.37   & 0.05    & 17.52 & 0.03 & 15.63 & 0.03 & 1 & 14.479 & 0.040 & 13.745 & 0.047 & 13.320 & 0.039 \\
15270276$+$4345172 & \nodata & \nodata & 16.89 & 0.03 & 15.08 & 0.03 & 1 & 14.002 & 0.026 & 13.310 & 0.026 & 13.021 & 0.028 \\
15480927$+$3227247 & 18.81   & 0.15    & 17.19 & 0.10 & 15.50 & 0.10 & 3 & 14.495 & 0.026 & 13.752 & 0.032 & 13.464 & 0.032 \\
15523734$+$2927591 & 17.97   & 0.05    & 16.30 & 0.03 & 14.73 & 0.03 & 1 & 13.797 & 0.029 & 13.201 & 0.035 & 12.881 & 0.033 \\
16223288$+$4237538 & 17.68   & 0.05    & 16.24 & 0.03 & 14.46 & 0.03 & 1 & 13.288 & 0.021 & 12.498 & 0.021 & 11.899 & 0.016 \\
16313278$+$3553285 & 19.89   & 0.15    & 18.13 & 0.10 & 16.38 & 0.10 & 3 & 15.466 & 0.050 & 14.846 & 0.060 & 14.326 & 0.062 \\
21051653$+$2514486 & 19.05   & 0.05    & 17.42 & 0.03 & 15.63 & 0.03 & 1 & 14.481 & 0.033 & 13.768 & 0.030 & 13.191 & 0.036 \\
21493784$-$1138285 & 17.66   & 0.05    & 15.83 & 0.03 & 14.04 & 0.03 & 1 & 13.002 & 0.023 & 12.262 & 0.023 & 11.898 & 0.026 \\
\enddata
\tablecomments{Table~\ref{tab:photometry} is published in machine-readable format.} 
\tablenotetext{a}{
Source of $BVI$ photometry. (1) $BVI$ from CCD measures on the NOFS 1.0 m telescope with estimated errors inflated to allow for significant differences between the spectral
energy distributions of dwarf carbon stars and the photometric standard stars employed. (2) $BVI$ from 17 independent measures on the NOFS 1.55 m telescope
employing a single channel photoelectric photometer with instrumental response bandpasses well matched to the standard system. (3) $BVI$ from SDSS $gri$
data transformed to the Johnson-Cousins system via the relations presented by \citet[][Table~7]{2007AJ....134..973I} and estimated errors inflated
to allow for significant instrument bandpass differences.}
\end{deluxetable*}

\subsection{Binary Orbits}
\label{sec:orbits}
\begin{figure}
\plotone{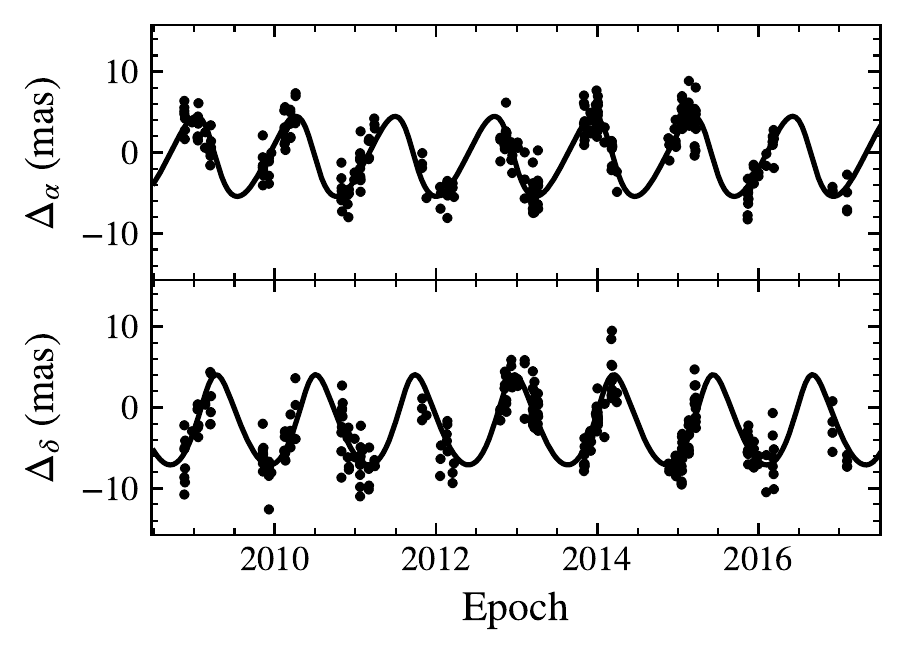}
\caption{Positions in right ascension and declination for LSPM J0742+4659 after
  removing the parallactic and proper motions.  The curves show the
  orbital-motion fit to these residuals as represented by the orbital
elements given in Table~\ref{tab:orbits}.}
\label{fig:orbit-1}
\end{figure}

\begin{figure}
\plotone{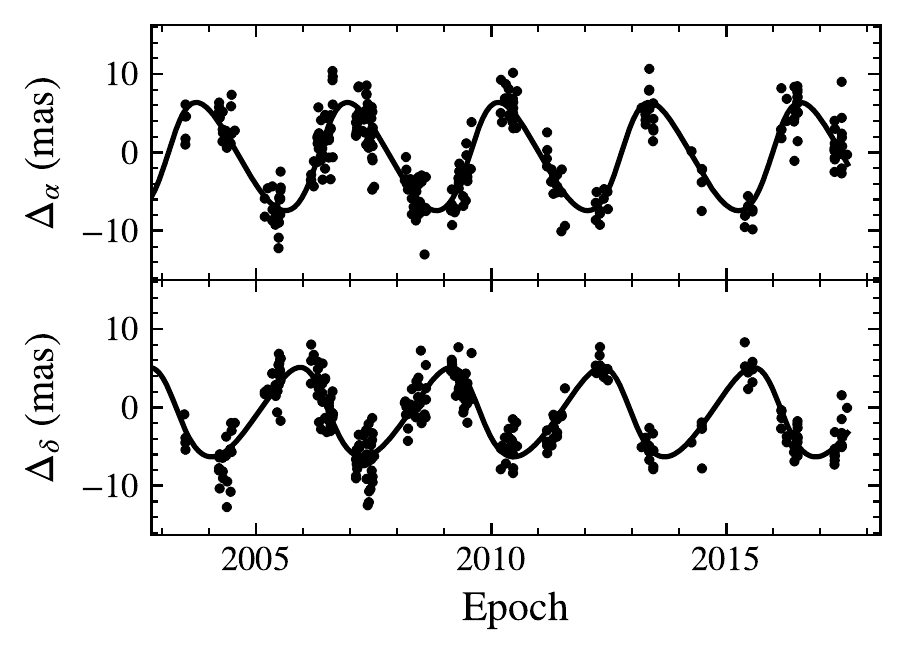}
\caption{Same as Figure~\ref{fig:orbit-1}, for LP225-12.}
\label{fig:orbit-2}
\end{figure}

\begin{figure}
\plotone{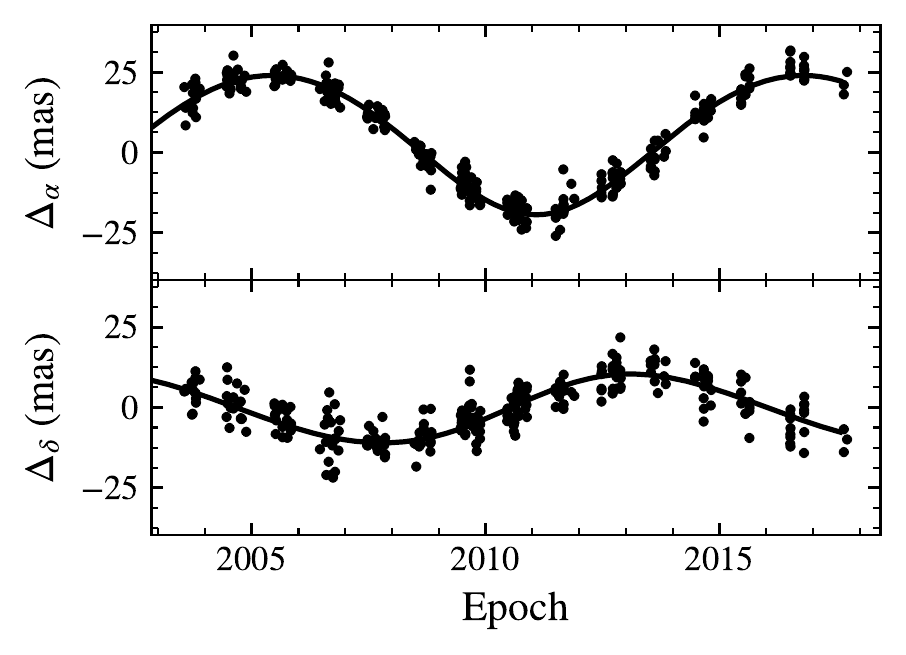}
\caption{Same as Figure~\ref{fig:orbit-1}, for LP758-43.}
\label{fig:orbit-3}
\end{figure}

Three stars in this sample, LSPM J0742+4659, LP225-12, and LP758-43, were found to have significant systematic
astrometric residuals from the default solution for parallax and
proper motion, indicating a periodic perturbation from an unseen
binary companion.  A solution for the orbit of the photocenter
(iterating between the parallax solution with the orbit removed, and
the orbit solution with the parallax and proper motion removed) was
carried out.  The results for the parallax and proper motion are given
in Table~\ref{tab:astrometry}, and for the orbital motion are given
in Table~\ref{tab:orbits} and plotted in Figures~\ref{fig:orbit-1} through
\ref{fig:orbit-3}.
The eccentricity for all three orbits is quite small and, within the
observational errors, is consistent with zero.
A fourth target G77-61 has been found to be a spectroscopic binary
\citep{1986ApJ...300..314D}, but we have not yet detected any
significant signature of the orbit in the astrometric data.
Another example of a dC spectrosopic binary
was found among the SDSS carbon stars \citep{2018ApJ...856L...2M}.
It is notable for its short orbital period of 2.9 d,
much shorter than \replaced{any of the four known dC binaries}{that of G77-61 and the three new systems reported here}.
This short period implies a different orbital
evolution during the mass-transfer process.
Other spectroscopic binaries are being found in a
new survey of dC stars \citep{whi18}.

\begin{deluxetable*}{lr@{ $\pm$ }lr@{ $\pm$ }lr@{ $\pm$ }l}
\tabletypesize{\small}
\tablewidth{0pt}
%%\tablenum{3}
\tablecaption{Orbits\added{ of New Astrometric dC Binaries}
\label{tab:orbits}}
\tablehead{
	\colhead{Parameter} &
	\multicolumn{2}{c}{LSPM J0742+4659} &
	\multicolumn{2}{c}{LP225-12} &
	\multicolumn{2}{c}{LP758-43}}
\startdata
Period (yr)     & 1.23   & 0.01 & 3.21   & 0.01 & 11.35  & 0.15 \\
$\alpha$\tablenotemark{a} (mas)  & 5.83   & 0.54 & 14.21  & 0.41 & 22.09  & 0.14 \\
i (deg)         & 150    & 4    & 72     & 2    & 63.3   & 1.0 \\
e               & 0.3    & 0.2  & 0.3    & 0.2  & 0.1    & 0.1 \\
$\omega$ (deg)  & 132    & 20   & 270    & 15   & 350    & 20 \\
$\Omega$ (deg)  & 154    & 5    & 128.5  & 0.5  & 282    & 3 \\
$T_0$           & 2014.1 & 0.1  & 2006.5 & 0.5  & 2011.0 & 0.5
\enddata
\tablenotetext{a}{Semimajor axis of the photocentric orbit.}
\end{deluxetable*}

%% import plots
%% plots.residuals('sdc0742+46c2')

%% import plots
%% plots.residuals('2m1622+42b5')

%% import plots
%% plots.residuals('lp758-43e2')

\section{Discussion}
\label{sec:discussion}
Studies by McClure and others \citep{1990ApJ...352..709M,
1997PASP..109..536M} have shown that the subgiant CH, giant CH, and
giant Ba stars are all binaries, most with periods 400 -- 4000 days and with
eccentricities smaller than for normal binaries.  These facts indicate
a previous phase of orbital dissipation during mass transfer from an
AGB companion, likely during wind accretion from the companion.

The orbits of the three dC binaries in \replaced{the}{this} paper are consistent with this
picture.  The three binaries have periods 450--4100 days, and their
eccentricites are small.  Using the parallaxes in Table~\ref{tab:astrometry},
the amplitudes of the photocenter orbits in Table~\ref{tab:orbits} are
0.9, 1.6, and 3.1 AU.  The true orbits will be larger
if the secondary star is contributing significant light to the
photocenter.  Because the data are taken with a red filter
(the wide-R A2-1 filter for LP758-43, and I-2 for the
other two binaries), and the companions are likely to be
white dwarfs, we expect that not much light is contributed
by the companions.  These amplitudes are comparable to those
for the CH and Ba stars studied by McClure, and similar to
the maximum size of the AGB star thought to be the source of the
carbon-rich material transfered to the dC star we see now.

The binaries found in this paper are expected to be
those with long periods among nearby targets,
because they will have the largest apparent (angular)
orbits that are easiest to detect astrometrically.
It is likely that more binaries with smaller astrometric
amplitudes are undetected in our sample.  Indeed, the
high frequency of binaries being found using radial
velocites \citep{whi18} indicates that most or all
dC stars are now in binaries with white dwarf companions.
Further study, and the determination of their orbits,
can confirm this expectation.

\begin{figure}
\plotone{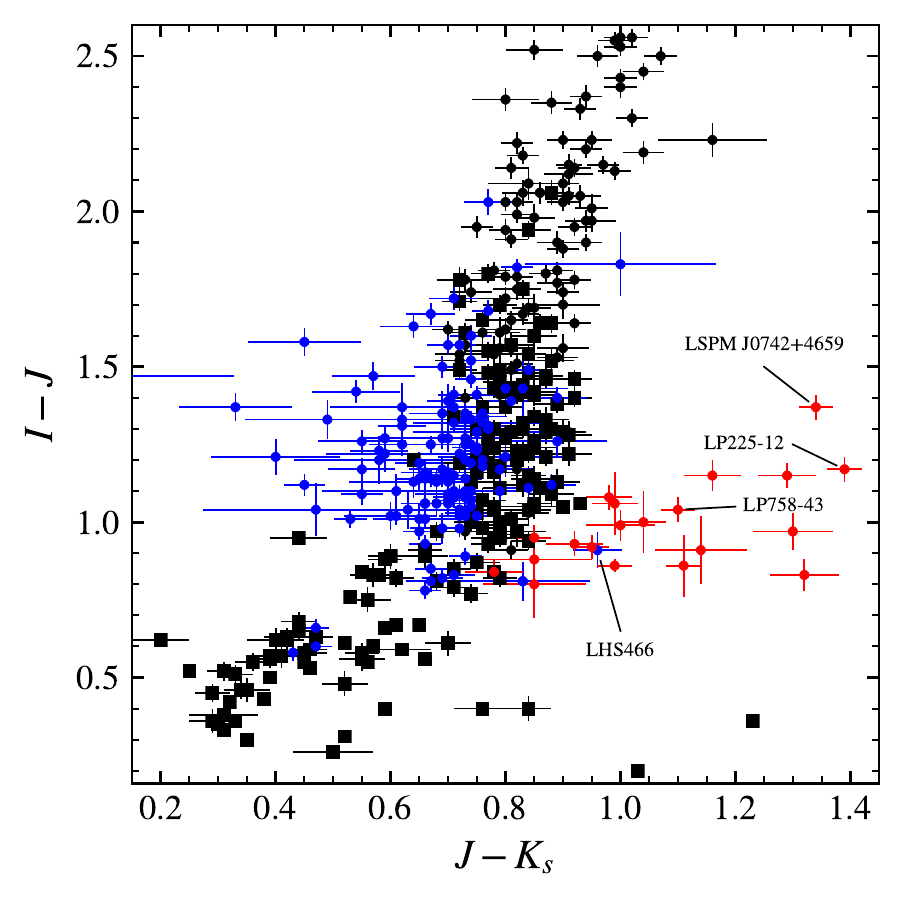}
\caption{$I-J$ versus $J-K_S$ for the dwarf carbon stars in this paper (red circles), along with late K and M dwarf (black circles) and subdwarf stars (blue circles) from
\citet{2017AJ....154..147D}, as well as selected late dwarf stars with parallaxes from {\it Hipparcos} and good $VI$ and 2MASS photometry (black squares).} 
\label{fig:IJ-JK}
\end{figure}

Figure~\ref{fig:IJ-JK} shows that the stars in this sample have $J-K_S$ colors
redder than oxygen-rich dwarfs and subdwarfs.  Presumably, their
$J$-band flux is suppressed by CN absorption, although some
contribution from $K_S$-band brightening from dust emission
in a few stars is possible.  The subdwarf LHS466, taken from
\citet{2017AJ....154..147D}, has $J-K_S$ = 0.88 measured by 2MASS, suggesting
it might be the brightest known dC star.  However, its UKIDSS
$J-K_S = 0.64$, corresponding to 0.66 on the 2MASS system, indicates
the 2MASS value is spurious or contaminated.  Therefore, Figure~\ref{fig:IJ-JK}
appears to provide a nearly clean separation of our dC stars.

%% import plots
%% plots.fig4()

Using $J-K_S$ colors to separate dC stars from the far more common
oxygen-rich dwarfs may provide an important means of identifying
a complete sample of dC stars in the solar neighborhood.
This step will be necessary to avoid the selection effects
in the present discovery of dC stars from SDSS, and thus in
the sample in this paper.  Therefore, Figure~\ref{fig:IJ-JK} may provide a
procedure to utilize Gaia data for an expanded sample
to facilitate understanding the true range of properties
of these stars.

\begin{figure}
\plotone{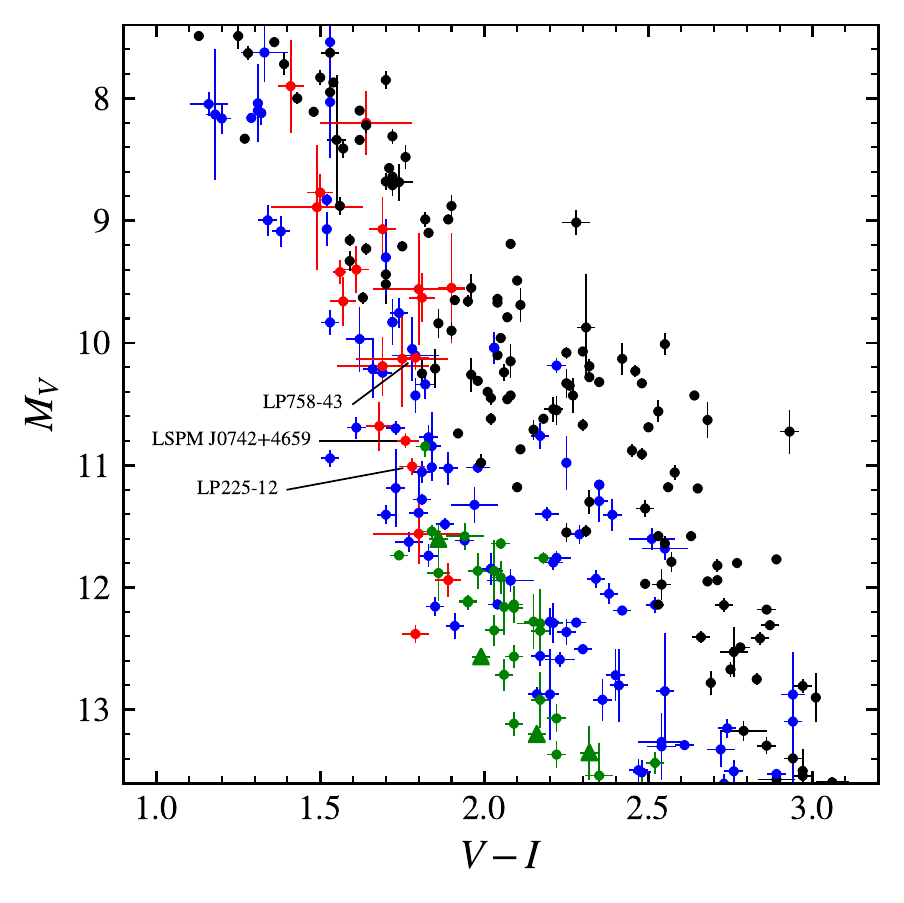}
\caption{$M_V$ versus $V-I$ for the dwarf carbon stars in this paper (red circles).  \added{The three new astrometric dC binaries are noted.}  Additional data points include late K and M dwarf (black circles) and subdwarf stars (blue circles) from \citet{2017AJ....154..147D} and {\it Hipparcos}.  Also plotted are extreme subdwarf (green circles) and ultra subdwarf (green triangle) M stars from \citet{2017AJ....154..147D}.} 
\label{fig:Mv-VI}
\end{figure}

Figure~\ref{fig:Mv-VI} shows a color-absolute magnitude diagram of the stars in
this paper.  They cover a range of $M_V$ of 7.9--12.4, and are subluminous by
up to 3 mag in $M_V$.  \added{We see that the close agreement of $M_V$ for the first 3 dCs with parallaxes \citep{1998ApJ...502..437H} was apparently fortuitous.}  Using the degree of subluminosity in
Figure~\ref{fig:Mv-VI}, together with their tangential velocities from
Table~\ref{tab:astrometry}, we can classify each star as probable disk or halo (Column~17 in Table~\ref{tab:astrometry}).
\footnote{
  Neither the tangential velocity nor the ``subluminosity'' in
  Figure~\ref{fig:Mv-VI}
are necessarily reliable indicators of the population of a star.
Halo stars can by chance have a small tangential velocity,
and disk stars can acquire high velocities through dynamical
interactions.  The high-velocity dC star found by \citet{2016ApJ...833..232P}
is discussed as possibly having acquired its high
velocity by ejection from a binary.
\replaced{Their}{Stars} positions in Figure~\ref{fig:Mv-VI} are affected by the
distorted spectral energy distribution of dC stars, most importantly by
absorption in $V$ by the C$_2$ Swan bands that shift $V-I$
toward a redder color for stars with strong bands.  A diagram
plotting luminosity versus temperature would be preferable.}
We find four stars are probable disk, 10 stars are probable halo,
two stars have intermediate values of subluminosity and velocity,
and four stars have contradictory values.  This result confirms
that the dC stars do have a wide range of properties, with
origins in the disk as well as the halo\added{\citep{2018MNRAS.tmp..864F}}.

\added{The first dwarf carbon star (G77-61) was identified as a dwarf from its parallax as measured by the USNO plate parallax program \citep{1977ApJ...216..757D}.  40 years later, parallaxes have been measured for only 20 additional dwarf carbon stars, all by the USNO CCD parallax program, including the 20 presented in this paper.}  Gaia DR2 will probably have parallax data for most of the stars in
this paper \citep{2017sf2a.conf..259K}.  The faintest stars may have
$G > 20$, fainter than the limit for DR2, and a few are likely to have
a binary perturbation detected by Gaia, and will be omitted from DR2
until a later data release.  For the stars that are in DR2, the formal
error of the DR2 parallaxes is estimated to be 0.1 -- 0.7 mas for the
magnitude range of our dC sample.  Therefore, DR2 data probably will
improve on the conclusions of this paper, other than the topic of
binary perturbations that DR2 will not address.  Later data releases
from Gaia will certainly improve on the distances, the binary
characteristics, and the sample size available to understand dC stars.

\acknowledgements
This publication
makes use of data products from the Two Micron All Sky Survey, which
is a joint project of the University of Massachusetts and the Infrared
Processing and Analysis Center/California Institute of Technology,
funded by the National Aeronautics and Space Administration and the
National Science Foundation.

\bibliography{bib}{}
\facility{USNO:61in (TEK2K, EEV24), USNO:40in, Sloan}

\end{document}